# Deep Learning from Dual-Energy Information for Whole-Heart Segmentation

# in Dual-Energy and Single-Energy Non-Contrast-Enhanced Cardiac CT

*Running title: Whole-Heart Segmentation in Non-Contrast CT*

**Steffen Bruns[1,2,3], Jelmer M. Wolterink[1,2,3], Richard A. P. Takx[4], Robbert W. van Hamersvelt[4],**

**Dominika Suchá[4], Max A. Viergever[2], Tim Leiner[4], Ivana Išgum[1,2,3,5]**

1 Department of Biomedical Engineering and Physics, Amsterdam UMC – location AMC, University of Amsterdam, Amsterdam, Netherlands

2 Image Sciences Institute, University Medical Center Utrecht, Utrecht, Netherlands

3 Amsterdam Cardiovascular Sciences, Amsterdam UMC, Amsterdam, Netherlands

4 Department of Radiology, University Medical Center Utrecht, Utrecht, Netherlands

5 Department of Radiology and Nuclear Medicine, Amsterdam UMC – location AMC, Amsterdam, Netherlands

Corresponding author:

Steffen Bruns (s.bruns@amsterdamumc.nl)

Department of Biomedical Engineering and Physics, Amsterdam UMC, University of Amsterdam, Meibergdreef 9, 1105 AZ Amsterdam, Netherlands

# Abstract

## Purpose

Deep learning-based whole-heart segmentation in coronary CT angiography (CCTA) allows the extraction of quantitative imaging measures for cardiovascular risk prediction. Automatic extraction of these measures in patients undergoing only non-contrast-enhanced CT (NCCT) scanning would be valuable, but defining a manual reference standard that would allow training a deep learning-based method for whole-heart segmentation in NCCT is challenging, if not impossible. In this work, we leverage dual-energy information provided by a dual-layer detector CT scanner to obtain a reference standard in virtual non-contrast (VNC) CT images mimicking NCCT images, and train a 3D convolutional neural network (CNN) for the segmentation of VNC as well as NCCT images.

## Methods

Eighteen patients were scanned with and without contrast enhancement on a dual-layer detector CT scanner. Contrast-enhanced acquisitions were reconstructed into a CCTA and a perfectly aligned VNC image. In each CCTA image, manual reference segmentations of the left ventricular (LV) myocardium, LV cavity, right ventricle, left atrium, right atrium, ascending aorta, and pulmonary artery trunk were obtained and propagated to the corresponding VNC image. These VNC images and reference segmentations were used to train 3D CNNs in a six-fold cross-validation for automatic segmentation in either VNC images or NCCT images reconstructed from the non-contrast-enhanced acquisition. Automatic segmentation in VNC images was evaluated using the Dice similarity coefficient (DSC) and average symmetric surface distance (ASSD). Automatically determined volumes of the cardiac chambers and LV myocardium in NCCT were compared to reference volumes of the same patient in CCTA by Bland-Altman analysis. An additional independent multi-vendor multi-center set of single-energy NCCT images from 290 patients was used for qualitative analysis, in which two observers graded segmentations on a five-point scale.

## Results

Automatic segmentations in VNC images showed good agreement with reference segmentations, with an average DSC of 0.897 ± 0.034 and an average ASSD of 1.42 ± 0.45 mm. Volume differences [95% confidence interval] between automatic NCCT and reference CCTA segmentations were -19 [-67; 30] mL for LV myocardium, -25 [-78; 29] mL for LV cavity, -29 [-73; 14] mL for right ventricle, -20 [-62; 21] mL for left atrium, and -19 [-73; 34] mL for right atrium, respectively. In 214 (74%) NCCT images from the independent multi-vendor multi-center set, both observers agreed that the automatic segmentation was mostly accurate (grade 3) or better.

## Conclusion

Our automatic method produced accurate whole-heart segmentations in NCCT images using a CNN trained with VNC images from a dual-layer detector CT scanner. This method might enable quantification of additional cardiac measures from NCCT images for improved cardiovascular risk prediction.

*Keywords*:     whole-heart segmentation, non-contrast-enhanced CT, convolutional neural network, deep learning, cardiac CT

## Introduction

Volumetric measurement of the cardiac substructures or *whole-heart segmentation* can provide valuable quantitative imaging measures for cardiovascular risk prediction[1-5]. Whole-heart segmentation can be performed in coronary CT angiography (CCTA)[6]. Since manual whole-heart segmentation in CCTA is time-consuming and tedious, a number of automatic methods have been developed for CCTA segmentation, including atlas-based[7] and machine learning-based methods[8-10].

CCTA images are often acquired in conjunction with non-contrast-enhanced CT (NCCT) images. While the CCTA is used to identify non-calcified atherosclerotic plaque and coronary stenoses[11], coronary artery calcification is quantified in NCCT[12]. In many cases where the NCCT image shows no coronary calcium, CCTA imaging will be omitted to avoid unnecessary radiation exposure and use of potentially nephrotoxic contrast agents[13]. However, quantification of the cardiac chambers and great vessels in these patients could provide information beyond the calcium score. For this purpose, it would be valuable to perform automatic whole-heart segmentation in NCCT images. Automatic methods for CCTA segmentation are not readily applicable to NCCT images as the contrast between tissues in these images is very different[14]. Moreover, retraining of these methods using manual reference segmentations in NCCT would be very challenging, as reference segmentations can hardly be obtained in NCCT due to the poor contrast between different cardiac tissues. Therefore, several works using reference segmentations obtained in NCCT have only focused on the aorta or the full heart, and not on cardiac chambers[15-17].

Studies aiming to segment cardiac chambers in NCCT either rely on ambiguous reference segmentations obtained in NCCT or challenging and error-prone inter-modality registrations to transfer manual reference segmentations from other modalities in which cardiac structures are clearly distinguishable. Kaderka et al.[18] directly obtained manual segmentations on NCCT images to build an atlas for the automatic segmentation of the cardiac chambers. Finnegan et al.[19] and Zhou et al.[20] also used a multi-atlas approach and combined manual segmentations of three and eight independent observers, respectively. The latter approach obtained the reference segmentations in NCCT while using registered CCTA images as a visual reference. Haq et al.[21] also used manual reference segmentations on NCCT images and trained a 2D CNN to segment the cardiac chambers. Although these approaches might reduce the segmentation ambiguity, the lack of contrast between cardiac structures makes it difficult to obtain reliable reference segmentations in NCCT directly. Shahzad et al.[22] propagated reference segmentations obtained in CCTA to corresponding NCCT images. However, patient and heart motion between acquisitions cause misalignment between CCTA and NCCT, and this approach required a challenging and error-prone inter-modality registration. Morris et al.[23] obtained reference segmentations jointly on registered pairs of NCCT and MRI images. All of the aforementioned methods might suffer from inaccurate alignment of NCCT images and reference segmentations, due to either ambiguous manual annotation in NCCT or errors in label propagation between modalities.

In this work, we exploit characteristics of a dual-layer detector CT scanner that allows the reconstruction of virtual non-contrast (VNC) images. VNC images mimic NCCT images and have shown to provide similar attenuation values to NCCT[24,25]. From a single acquisition with contrast enhancement on a dual-layer detector CT scanner, both a CCTA and a perfectly aligned VNC image can be reconstructed. We perform manual annotation in CCTA images with good visibility of borders between cardiac structures. As the VNC image is perfectly aligned with the CCTA image, reference segmentations obtained on CCTA directly delineate structures in VNC. This enables us to omit challenging and error-prone inter-modality registration. VNC images and reference segmentations are used to train a 3D convolutional neural network (CNN) for segmentation in VNC as well as NCCT images.

We evaluate segmentation performance on left ventricular (LV) myocardium, LV cavity, right ventricle (RV), left atrium (LA), right atrium (RA), ascending aorta, and pulmonary artery trunk in both VNC and NCCT images in a six-fold cross-validation experiment. Moreover, we qualitatively evaluate automatic segmentation on NCCT images in a multi-vendor multi-center dataset containing both thin- and thick-slice NCCT images using ratings by two expert observers.

## Materials and Methods

### Datasets

Two datasets were used in this work: First, a set of non-contrast-enhanced and contrast-enhanced acquisitions on a dual-layer detector CT scanner which we refer to as *dual-energy set*. Second, an independent test set of non-contrast-enhanced and contrast-enhanced acquisitions on four different single-energy CT scanners which we refer to as *single-energy set*.

The dual-energy set contained CT images of 18 patients that were scanned using a dual-layer detector CT scanner (IQon, Philips Healthcare, Best, The Netherlands) in an ongoing prospective study aimed at assessing the diagnostic performance of dual-layer detector CT for the identification of functionally significant coronary artery stenosis[26]. Each patient received a non-contrast-enhanced acquisition which was reconstructed into an NCCT image. This was followed by an acquisition with contrast enhancement. By using the dual-energy information, the contrast-enhanced acquisition could be reconstructed into both a CCTA image and a VNC image (Fig. 1).

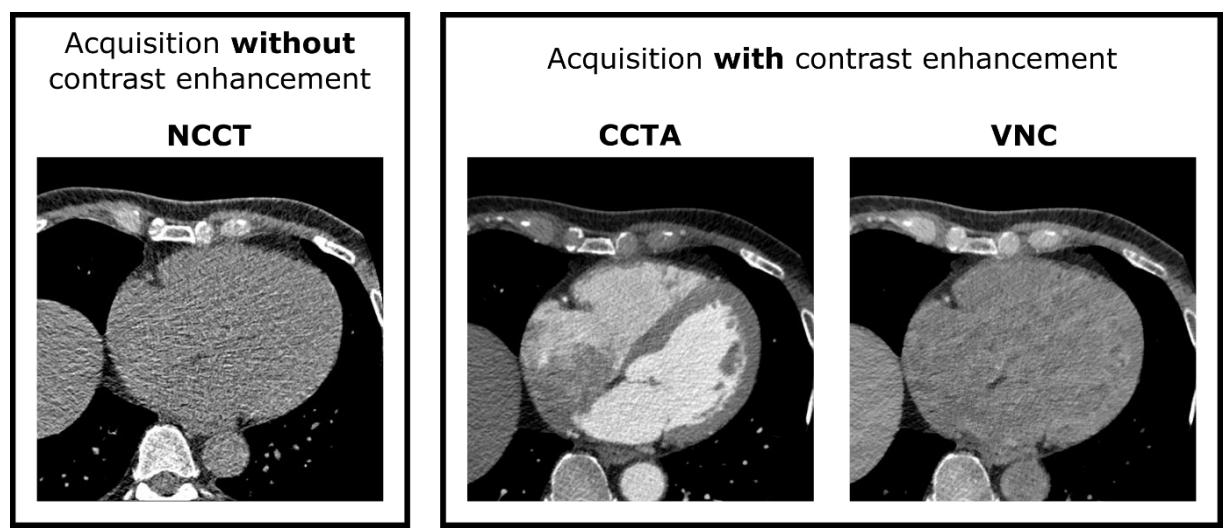

*Figure 1: Different CT images of the same patient, acquired on a dual-layer detector CT scanner. Axial plane of non-contrast-enhanced CT (NCCT) image reconstructed from an acquisition without contrast enhancement. Axial planes of a coronary CT angiography (CCTA) image and a virtual non-contrast (VNC) image reconstructed from an acquisition with contrast enhancement.*

To investigate the extent to which VNC images mimic NCCT images, we performed an equivalence test in the dual-energy set. In line with the existing literature[24], we manually selected one square region of interest (ROI) of 1 cm$^2$ at the aortic root in each CCTA, VNC, and NCCT image and computed the mean attenuation in this ROI. Attenuation values were substantially higher in CCTA (median 324, interquartile range [IQR] 299-345 HU) than in VNC (median 37, IQR 32-42 HU) and NCCT (median 43, IQR 38-49 HU) (Fig. 2). The mean difference between attenuation values in the ROIs in VNC and NCCT images was <15 Hounsfield units (HU) in 13/18 (72.2%) exams, <10 HU in 9/18 (50%) exams, and <5 HU in 7/18 (38.9%) exams. Hence, the mean attenuation values between VNC and NCCT were equivalent with a threshold of 15 HU difference (under the significance level of 0.05), thus below the mean standard deviation of 28 HU in the selected ROIs in NCCT. While this shows that VNC and NCCT images have similar mean attenuation values, visual inspection of the VNC images revealed that virtual contrast removal was not perfect. As illustrated in Figure 1, VNC images contained residual contrast patterns, and noise levels in NCCT images appeared to be higher.

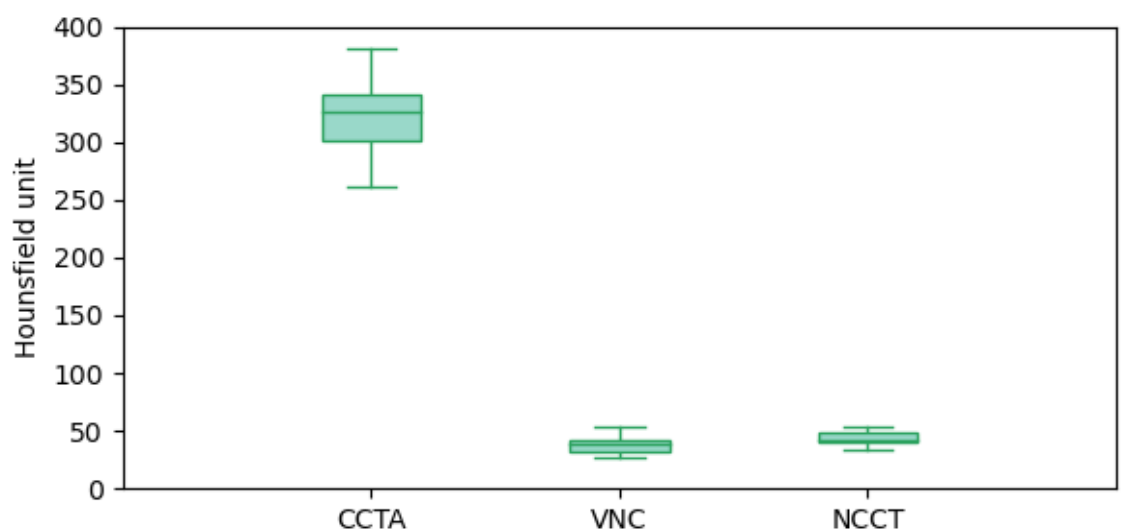

*Figure 2: Mean attenuation values in 1 cm$^2$ ROIs at the aortic root in coronary CT angiography (CCTA), virtual non-contrast (VNC), and non-contrast-enhanced CT (NCCT) images from the dual-energy set.*

For independent testing on a large set of commonly available single-energy CT images, we combined CT images from two different sources into the single energy set. First, 218 patients that were scanned using a conventional single-energy CT scanner (Brilliance iCT, Philips Healthcare, Best, The Netherlands) during routine clinical care. Second, to also evaluate our method on images from different vendors at different centers, we included CT images of 72 patients from the orCaScore[27] study. Those CT examinations were carried out in four different medical centers with four different single-energy CT scanners (Brilliance iCT, Philips Healthcare, Best, The Netherlands; Lightspeed VCT, General Electric, Milwaukee, USA; SOMATOM Definition Flash, Siemens Healthineers, Erlangen, Germany; Aquilion One, Toshiba Medical Systems, Otawara, Japan). As in the dual-energy set, each patient in the single-energy set was scanned without and with contrast administration. However, because these images were acquired on single-energy CT scanners, no VNC image could be reconstructed for these patients. Further specifications of data acquisition and reconstruction are listed in Table 1.

*Table 1: Overview of datasets, split into dual-energy set and single-energy set. NCCT = non-contrast-enhanced CT, CCTA = coronary CT angiography, VNC = virtual non-contrast.*

| Specification | Dual-energy set | Single-energy set | |
|---|---|---|---|
| Source | CLARITY study[26] | Routine clinical care | orCaScore study[27] |
| Date range | 2017 – 2018 | 2012 – 2013 | 2011 – 2014 |
| Hospital | University Medical Center Utrecht, NL | University Medical Center Utrecht, NL | University Medical Center Utrecht, NL; Antwerp University Hospital, BE; Radboud University Medical Center, NL; University Medical Center Groningen, NL |
| Scanner(s) | Philips IQon | Philips Brilliance iCT | Philips Brilliance iCT, General Electric Lightspeed VCT, Siemens SOMATOM Definition Flash, Toshiba Aquilion One |
| # Patients | 18 (16 M) | 218 (169 M) | 72 (36 M) |

| Age | 62.3 ± 8.5 years | 51.4 ± 12.4 years | 58.7 ± 8.8 years |
|---|---|---|---|
| Tube voltage | 120 kVp | 120 kVp | 80, 100, 120 kVp |
| Tube current | 441 mAs | 10 − 993 mAs | 130 − 993 mAs |
| ECG-triggering phase | 78% | 78% | 70 − 81% |
| In-plane resolution | 0.346 − 0.5 mm | 0.287 − 0.488 mm | 0.314 − 0.625 mm |
| Slice thickness NCCT | 0.9 mm | 0.9 mm | 2.5 − 3.0 mm |
| Slice thickness CCTA | 0.9 mm | 0.9 mm | 0.5 − 0.9 mm |
| Reconstructed images | 18 NCCT<br>18 CCTA<br>18 VNC | 218 NCCT<br>218 CCTA | 72 NCCT<br>72 CCTA |

## Reference Segmentations

Reference segmentations for seven cardiac structures were obtained in the 18 CCTA images of the dual-energy set. These structures were: LV myocardium, LV cavity, RV, LA, RA, ascending aorta, and pulmonary artery trunk until the first bifurcation. Initial reference segmentations were obtained using a previously described automatic method[28], and medical students subsequently performed manual voxel-wise correction using 3D Slicer (3D Slicer 4.8.1, http://www.slicer.org). All reference segmentations were verified and if necessary corrected by a resident (RAPT, with 7 years of experience in cardiothoracic imaging). No reference segmentations were obtained in NCCT images.

## Automatic Segmentation

Since the CCTA and VNC images were reconstructed from the same CT acquisition, the reference segmentations obtained on CCTA could directly be propagated to the VNC images. CNN training was thus performed using VNC images together with reference segmentations obtained on CCTA (Fig. 3). Once a CNN was trained, it could be used to segment the seven cardiac structures in both VNC and NCCT images. No manual segmentation on NCCT images was necessary and no image registration had to be performed.

**Training (dual-energy set)**

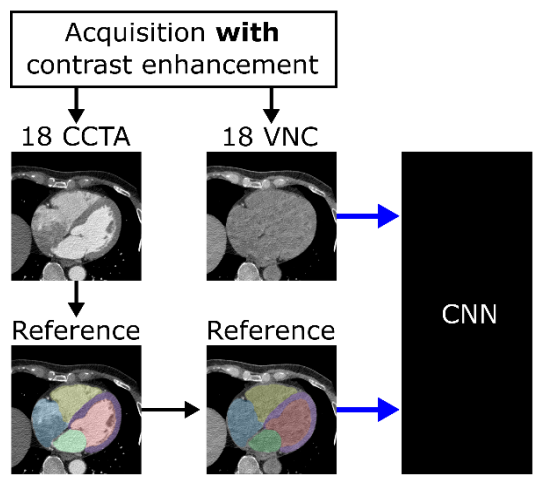

**Testing (dual-energy set)**                                    **Testing (single-energy set)**

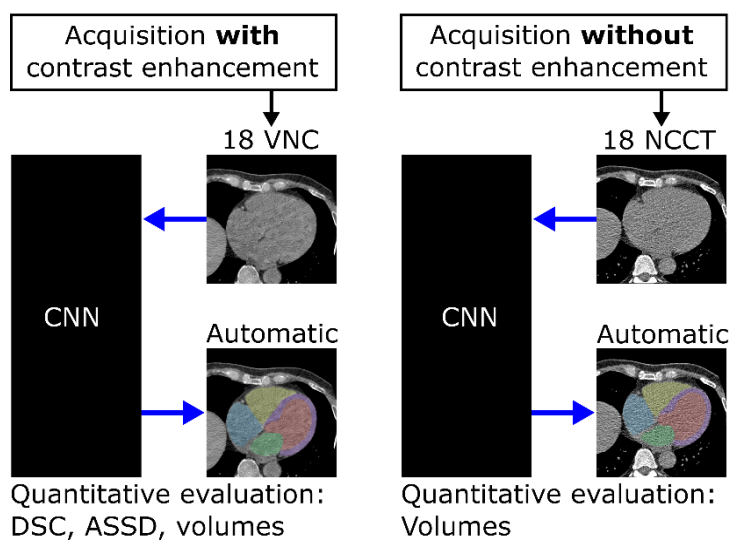    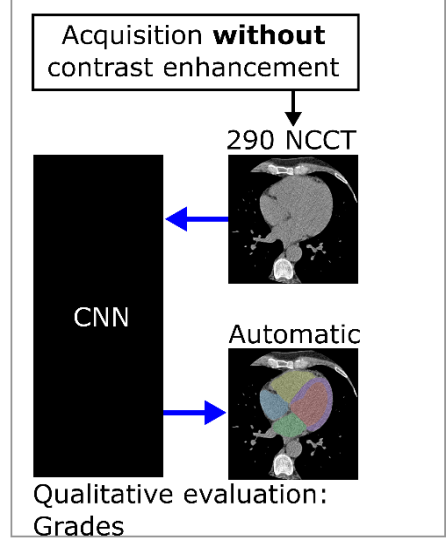

*Figure 3: Method overview. Coronary CT angiography (CCTA) and virtual non-contrast (VNC) images are reconstructed from the same acquisition with contrast enhancement on a dual-layer detector CT scanner. Reference segmentations are obtained on the CCTA images and transferred to VNC. A convolutional neural network (CNN) is trained to segment seven cardiac structures in VNC. The trained CNN can be used to segment VNC or non-contrast-enhanced CT (NCCT) images.*

The 3D CNN architecture used in this work is based on a previously proposed 2D architecture[29]. The CNN contains an encoding path with three downsampling layers with strided convolutions and a decoding path with three upsampling layers with transposed convolutions (Table 2). The encoding and decoding paths are connected by a series of six ResNetBlocks that operate on the downsampled input. We adapted the CNN to operate on 3D 256 x 256 x 5 voxel sub-image inputs. This allows the network to take sufficient contextual information into account in the axial direction and encourages smoothness between predictions across adjacent axial slices. After each but the last convolutional layer of the CNN, 3D batch normalization and rectified linear units were applied.

*Table 2: Details of the segmentation network architecture. The input image is downsampled through layers 1-4, transformed through layers 5-10, and upsampled through layers 11-14.*

| Layer | Type | Filter size | Stride | Channels | BatchNorm | Activation function |
|-------|------|-------------|--------|----------|-----------|---------------------|
| 1 | Conv | 5x5x5 | 1 | 16 | Yes | ReLU |
| 2 | Strided Conv | 3x3x1 | 2 | 32 | Yes | ReLU |
| 3 | Strided Conv | 3x3x1 | 2 | 64 | Yes | ReLU |
| 4 | Strided Conv | 3x3x1 | 2 | 128 | Yes | ReLU |
| 5-10 | ResNetBlock | 3x3x1 | 1 | 128 | Yes | ReLU |
|  |  | 3x3x1 | 1 | 128 | Yes |  |
| 11 | Transposed Conv | 3x3x1 | 2 | 64 | Yes | ReLU |
| 12 | Transposed Conv | 3x3x1 | 2 | 32 | Yes | ReLU |
| 13 | Transposed Conv | 3x3x1 | 2 | 16 | Yes | ReLU |
| 14 | Conv | 5x5x1 | 1 | 8 | No | Softmax |

To avoid potentially very high noise levels in NCCT, the images were smoothed with a 2D Gaussian filter ($\sigma$ = 2 mm) prior to our experiments. As a normalization, they were linearly rescaled between -1024 and 3071 Hounsfield units to a [0, 1] intensity range. To correct for image resolution differences, all images were resampled to an isotropic resolution of 0.8 x 0.8 x 0.8 mm$^3$. All CNNs were trained using mini-batches consisting of 32 randomly sampled 256 x 256 x 5 voxel sub-images. CNNs were trained for 10,000 iterations using Adam as the optimizer, an initial learning rate of 0.001, and a 70% learning rate decay every 2000 iterations. The loss function was defined as the negative sum of soft Dice similarity coefficients (DSC) over all classes[30]. Both the training loss and the validation loss were tracked to be sure that no severe overfitting occurs. To obtain predictions for all voxels in an image, the CNN was applied to 256 x 256 x 5 voxel sub-images and outputs were stitched together. To obtain only one contiguous segmentation for each structure, the output of the CNN was processed using a connected component labeling. For each structure, except for the pulmonary artery trunk which could consist of multiple components, we included only the largest connected component in the final segmentation result.

The dual-energy set was randomly split into six folds consisting of VNC images from three patients each. For each of the six folds, images of the three held-out patients were used for validation and testing and images of the remaining 15 patients were used for training. For each fold, an ensemble of three CNNs with different random initializations was trained. For each test image, the two remaining held-out images were used for validation by determining the stopping criterion for training based on the combined loss on these two images. Thus, each test image was segmented with a unique ensemble of three CNN instances in which the output probabilities were averaged. The corresponding NCCT images from the dual-energy set were segmented with the same ensembles of CNNs. The NCCT images from the independent test set, however, were segmented using an ensemble of all 18 CNNs trained in the cross-validation experiments, in which output probabilities were also averaged.

## Quantitative Evaluation

Reference segmentations were available for all CCTA images in the dual-energy set. This allowed quantitative evaluation of automatic segmentations obtained in the corresponding VNC images reconstructed from the same acquisition. For this, we used the DSC and average symmetric surface distance (ASSD). NCCT images were reconstructed from a different acquisition than CCTA and VNC images, and thus are not aligned with the reference segmentations obtained in these images. This precludes the evaluation of NCCT segmentations using DSC or ASSD metrics. Instead, automatically determined volumes of the structures with full coverage in both CT acquisitions were compared to those in the reference standard by Bland-Altman analysis and a two-paired t-test. These structures were the LV myocardium, LV cavity, RV, LA, and RA as well as the sum of these volumes, i.e. the full heart volume.

## Qualitative Evaluation

Since reference segmentations can hardly be obtained in NCCT images and no VNC images were available in the single-energy set, the automatic segmentations obtained in these images were evaluated qualitatively. Two residents (RAPT and DS, with 7 years of experience in cardiothoracic imaging) independently inspected the automatic segmentations as an overlay on the NCCT image, with the CCTA image of the same patient shown for reference. The observers assigned grades on a five-point scale (Table 3) to the automatic segmentations based on previously proposed criteria[31]. In this qualitative scale, deviations are defined as maximum distances, equivalent to the Hausdorff distance.

*Table 3: Five-point scale for qualitative evaluation of automatic segmentations, with definitions as proposed by Abadi et al.[31].*

| Grade | Description |
|-------|-------------|
| 1 | Very accurate segmentation with deviations up to 1 mm |
| 2 | Accurate segmentation where 1-2 structures deviate up to 3 mm |
| 3 | Mostly accurate segmentation where one structure deviates up to 1 cm or >2 structures deviate up to 3 mm |
| 4 | Segmentation where up to 50% of the target area is not segmented correctly |
| 5 | Failed segmentation where more than 50% of the target area is not segmented correctly |

The linearly weighted Cohen's Kappa score between both observers was computed on a confusion matrix representing their assessment. In addition, the observers were asked to indicate for which individual structures segmentations deviated more than 1 mm. The automatic segmentations obtained on NCCT images in the training set were evaluated in the same way by one of the observers.

# Results

## Quantitative Evaluation

Table 4 lists the DSC and ASSD between reference segmentations and automatic segmentations on VNC images obtained using cross-validation in the dual-energy set. Automatic segmentation of all cardiac chambers as well as the ascending aorta achieved a DSC >= 0.9 and an ASSD <= 2 mm. In terms of DSC and ASSD, performance was best on the ascending aorta likely because of its clear visibility even without contrast enhancement. While the DSC was lowest on the LV myocardium, the ASSD was worst on the pulmonary artery trunk. The border of the LV myocardium with the LV cavity is hardly visible and its shape makes it prone to drops in DSC in case of segmentation errors. For the pulmonary artery trunk, segmentation beyond the first bifurcation might cause a high ASSD in some cases.

*Table 4: Quantitative comparison of automatic and reference segmentations on VNC images in the dual-energy set. Mean ± standard deviation of Dice similarity coefficient (DSC) and average symmetric surface distance (ASSD).*

| Structure | DSC | ASSD |
|---|---|---|
| Left ventricular myocardium | 0.835 ± 0.049 | 1.18 ± 0.29 mm |
| Left ventricular blood cavity | 0.903 ± 0.029 | 1.72 ± 0.59 mm |
| Right ventricle | 0.916 ± 0.030 | 1.43 ± 0.58 mm |
| Left atrium | 0.921 ± 0.023 | 1.17 ± 0.43 mm |
| Right atrium | 0.906 ± 0.036 | 1.38 ± 0.50 mm |
| Ascending aorta | 0.940 ± 0.014 | 0.74 ± 0.27 mm |
| Pulmonary artery trunk | 0.859 ± 0.061 | 2.29 ± 1.88 mm |
| Average | 0.897 ± 0.034 | 1.42 ± 0.45 mm |

Table 5 lists a comparison between the proposed method and previously published studies on whole-heart segmentation in NCCT. Although a direct comparison is not possible as different data sets have been used in different studies, these results show that our automatic segmentation method is competitive with other methods in terms of DSC. Moreover, to the best of our knowledge, we are the first to segment LV cavity and LV myocardium as separate classes. Nevertheless, it should be noted that our results were obtained on VNC images instead of native NCCT images. We also compare the results in Table 5 with those obtained in CCTA images in a recent multi-modality whole-heart segmentation challenge[10]. The top-three performing methods in this challenge achieved an average DSC of 0.908 ± 0.086, 0.894 ± 0.030, and 0.890 ± 0.049, on CCTA images, respectively. This is comparable to the DSC scores obtained by the methods in Table 5.

*Table 5: Comparison of results with related works that automatically segment cardiac chambers in non-contrast-enhanced CT (NCCT). [a]average result on the four cardiac chambers. [b]average result on ascending aorta, pulmonary artery, and superior vena cava. [c]The complete data set consisted of a mix of NCCT and CCTA images. CCTA = coronary CT angiography.*

| | Structures | Method | Reference obtained in | # Train/ test | Dice similarity coefficient on | | | | | |
| --- | --- | --- | --- | --- | --- | --- | --- | --- | --- | --- |
| | | | | | Left ventricle | Right ventricle | Left atrium | Right atrium | Ascending aorta | Pulmo-nary artery |
| Kaderka et al.[18] | Chambers, left anterior descending artery | Multi-atlas | NCCT | 27/27 | 0.85 ± 0.04 | 0.79 ± 0.06 | 0.76 ± 0.06 | 0.76 ± 0.06 | - | - |
| Finnegan et al.[19] | Chambers, great vessels, coronary arteries, valves | Multi-atlas | NCCT | 20, cross-valida-tion | 0.862 ± 0.072 | 0.808 ± 0.053 | 0.852 ± 0.033 | 0.821 ± 0.054 | 0.796 ± 0.072 | 0.668 ± 0.094 |
| Zhou et al.[20] | Chambers, great vessels, coronary arteries | Multi-atlas | NCCT | 12/19 | 0.79 ± 0.13 | 0.70 ± 0.15 | 0.77 ± 0.07 | 0.72 ± 0.14 | - | - |
| Haq et al.[21] | Chambers, great vessels | 2D CNN | NCCT and CCTA[c] | 192/24 | **0.925** | 0.866 | 0.865 | 0.871 | 0.827 | **0.882** |
| Shahzad et al.[22] | Chambers, aorta | Multi-atlas | CCTA | 90, cross-valida-tion | 0.90 | 0.87 | 0.91 | **0.91** | 0.90 | - |
| Morris et al.[23] | Chambers, great vessels, coronary arteries | 3D CNN | MRI/ NCCT | 25/7 | 0.88 ± 0.03[a] | 0.88 ± 0.03[a] | 0.88 ± 0.03[a] | 0.88 ± 0.03[a] | 0.85 ± 0.03[b] | 0.85 ± 0.03[b] |
| Pro-posed | Chambers ascending aorta, pulmonary artery | 3D CNN | CCTA | 18, cross-valida-tion | 0.903 ± 0.029 | **0.916 ± 0.030** | **0.921 ± 0.023** | 0.906 ± 0.036 | **0.940 ± 0.014** | 0.859 ± 0.061 |

The volumes of automatically segmented structures in VNC images were not significantly different ($0.075 \leq p \leq 0.550$) from CCTA reference volumes with mean bias and 95% confidence intervals of -5 [-46; 36] mL for the LV myocardium, -3 [-42; 36] mL for the LV cavity, -4 [-22; 14] mL for the RV, -2 [-15; 11] mL for the LA, -3 [-25; 11] mL for the RA, and -17 [-93; 59] mL for the full heart. However, we found that automatically obtained absolute volumes in NCCT images were significantly lower (under the significance level of 0.05) than those in the reference standard derived from the CCTA (Fig. 4).

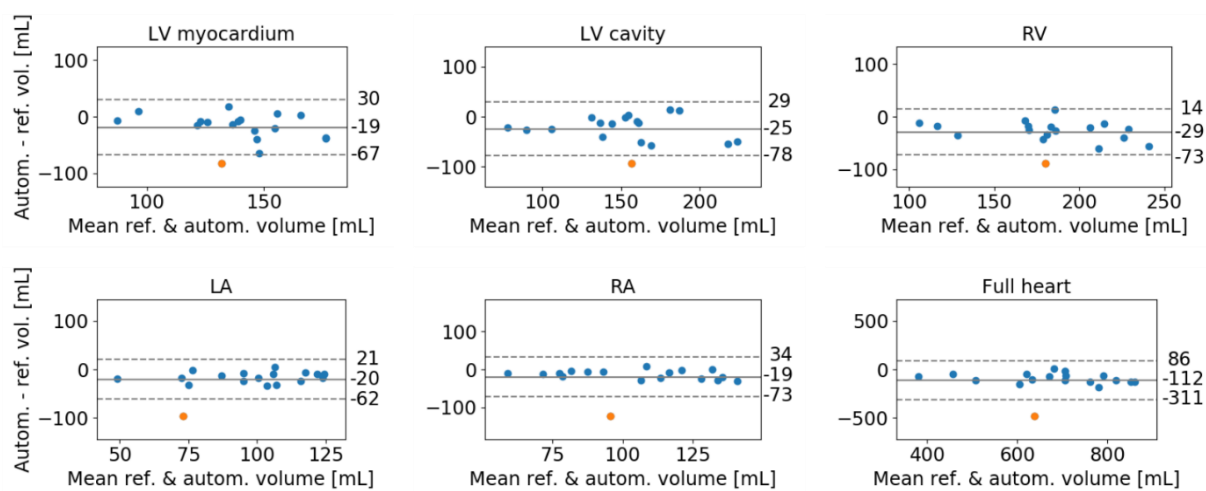

*Figure 4: Bland-Altman plots comparing the volumes obtained from automatic segmentations on non-contrast-enhanced CT images versus reference volumes from coronary CT angiography images in the dual-energy set. The full heart volume as the sum of the volumes of LV myocardium and the four heart chambers. The orange dot indicates an image with strong artifacts due to a prosthetic mitral valve. LV = left ventricular, RV = right ventricle, LA = left atrium, RA = right atrium*

Figure 5 shows an example for one patient in the dual-energy set. Despite the high visual quality of the automatic NCCT segmentation (graded as very accurate by the expert observer), absolute volumes of structures in this segmentation were around 20% lower than in CCTA. However, an additional evaluation showed that relative structure sizes within the heart were not significantly different (under the significance level of 0.05) in CCTA, VNC, and NCCT (Table 6).

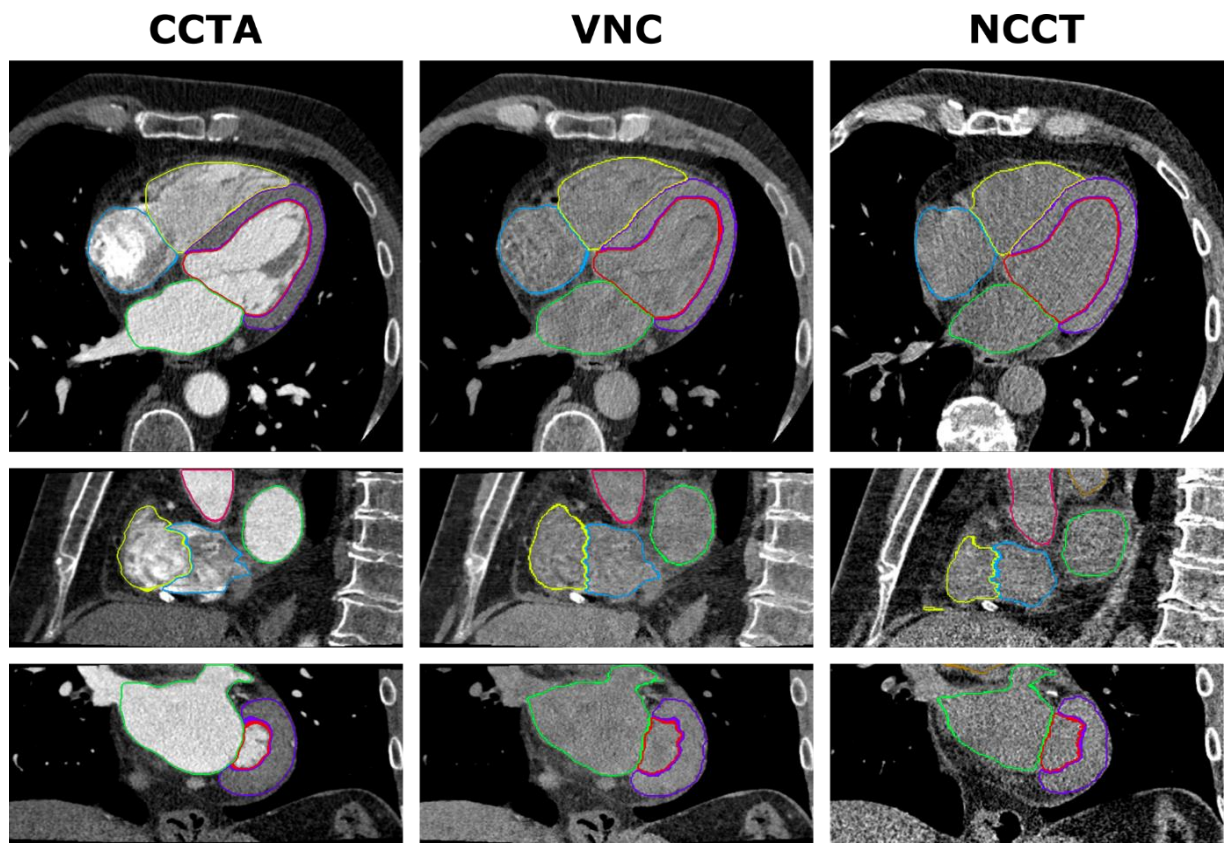

*Figure 5: Segmentations on axial (top), sagittal (middle) and coronal (bottom) views of a patient in the dual-energy set. Reference segmentation on coronary CT angiography image (CCTA) (left), automatic segmentation on virtual non-contrast image (VNC) (middle), and automatic segmentation on non-contrast-enhanced CT image (NCCT) (right).*

*Table 6: Absolute and relative volumes (relative to the respective full heart volume) derived from the reference segmentations on coronary CT angiography (CCTA), the automatic segmentations on virtual non-contrast (VNC), and automatic segmentations on non-contrast-enhanced CT (NCCT) images. Full heart volume as the sum of the volumes of the left ventricular myocardium and the cardiac chambers.*

| Structure | Reference (CCTA/VNC) | Automatic VNC | Automatic NCCT |
|---|---|---|---|
| Left ventricular myocardium | 148.7 ± 29.0 mL  20.3 ± 4.0% | 143.6 ± 22.9 mL  20.1 ± 3.2% | 129.8 ± 22.7 mL  20.9 ± 3.7% |
| Left ventricular blood cavity | 164.9 ± 41.4 mL  22.5 ± 5.7% | 162.0 ± 41.9 mL  22.7 ± 5.9% | 140.4 ± 37.7 mL  22.7 ± 6.0% |
| Right ventricle | 196.9 ± 40.6 mL  26.9 ± 5.5% | 192.9 ± 37.9 mL  27.0 ± 5.3% | 167.6 ± 34.9 mL  27.1 ± 5.6% |
| Left atrium | 107.4 ± 20.0 mL  14.7 ± 2.7% | 105.4 ± 21.6 mL  14.7 ± 3.0% | 87.0 ± 26.1 mL  14.0 ± 4.2% |
| Right atrium | 114.2 ± 28.2 mL  15.6 ± 3.9% | 110.9 ± 26.4 mL  15.5 ± 3.7% | 94.7 ± 27.3 mL  15.3 ± 4.4% |
| Full heart | 732.1 ± 141.6 mL  100.0 ± 19.9% | 714.9 ± 138.8 mL  100.0 ± 19.4% | 619.6 ± 133.8 mL  100.0 ± 21.6% |

## Qualitative Evaluation

Observer 1 assigned 129 of the 290 (44%) automatic segmentations on NCCT images to grade 1, 67 (23%) segmentations to grade 2, 39 (13%) segmentations to grade 3, 48 (17%) segmentations to grade 4, and 7 (2%) segmentations to grade 5. Observer 2 assigned 73 (25%), 83 (29%), 68 (23%), 64 (22%), and 2 (1%) automatic segmentations to grade 1 through 5, respectively. Table 7 shows a confusion matrix reporting the grades assigned to the segmentations by the two observers.

*Table 7: Confusion matrix showing the grades assigned to automatic segmentations on non-contrast-enhanced CT images in the single-energy set by the two observers. Grade 1 corresponds to very accurate segmentations, grade 5 to failed segmentations.*

| Obs 1 \ 2 | Grade 1 | Grade 2 | Grade 3 | Grade 4 | Grade 5 | Total |
|-----------|---------|---------|---------|---------|---------|-------|
| Grade 1   | 67      | 48      | 12      | 2       | 0       | 129   |
| Grade 2   | 6       | 28      | 29      | 4       | 0       | 67    |
| Grade 3   | 0       | 6       | 18      | 15      | 0       | 39    |
| Grade 4   | 0       | 1       | 9       | 38      | 0       | 48    |
| Grade 5   | 0       | 0       | 0       | 5       | 2       | 7     |
| Total     | 73      | 83      | 68      | 64      | 2       | 290   |

The two observers assigned a grade of 3 or better, i.e. mostly accurate to very accurate, to 235 (81%) and 224 (77%) NCCT images respectively, and agreed on this categorization in 214 (74%) cases. However, agreement on fine-grained five-class grading was lower at 53%, corresponding to a linearly weighted kappa of 0.59. Scores assigned by observer 1 were generally better (mean 2.09) than those assigned by observer 2 (mean 2.44). The mean grades per vendor on the thick-slice NCCT images from the orCaScore[27] study, averaged over both observers and 18 images per vendor, were all between 2.31 and 2.44, while the mean grade on the thin-slice NCCT images was 2.22.

Table 8 shows how often segmentations of individual cardiac structures deviated less than 1 mm according to both observers. These results show that both atria, the ascending aorta, and the pulmonary artery trunk were segmented very accurately in the majority of images, and the LV myocardium, the LV cavity, and the RV were more often affected by deviations of more than 1 mm.

*Table 8: Amount of automatic segmentations in which the individual structures are segmented with deviations of less than 1 mm according to the observers. LV = left ventricular, RV = right ventricle, LA = left atrium, RA = right atrium*

| Observer | LV myocardium | LV cavity | RV | LA | RA | Ascending aorta | Pulmonary artery trunk |
|---|---|---|---|---|---|---|---|
| Observer 1 | 115 (40%) | 171 (59%) | 128 (44%) | 229 (79%) | 225 (78%) | 248 (86%) | 258 (89%) |
| Observer 2 | 115 (40%) | 176 (61%) | 147 (51%) | 259 (89%) | 237 (82%) | 253 (87%) | 268 (92%) |
| Consensus | 90 (31%) | 135 (47%) | 110 (37%) | 223 (77%) | 212 (73%) | 238 (82%) | 251 (87%) |

Figure 6 shows very accurate automatic segmentations on NCCT images from the single-energy set. Two very accurate segmentations (grade 1) on thin-slice NCCT images are shown in Fig. 6(a) and 6(c),and two on thick-slice images in Fig. 6(b) and 6(d). Figure 7 shows automatic segmentations on NCCT images from the single-energy set on which the automatic method produced incorrect (grade 4) or failed (grade 5) segmentations. Figure 7(a) shows an incorrect segmentation (grade 4) due to a prosthetic mitral valve. In all three views, one can observe that the segmentation fails around the mitral valve while the segmentation appears accurate in regions further away from it. Figure 7(b) shows a failed segmentation (grade 5) due to a strong anatomical variation of the heart shape. Figures 7(c) and 7(d) show incorrect (grade 4) segmentations which could not clearly be linked to metal implants or anatomical variations.

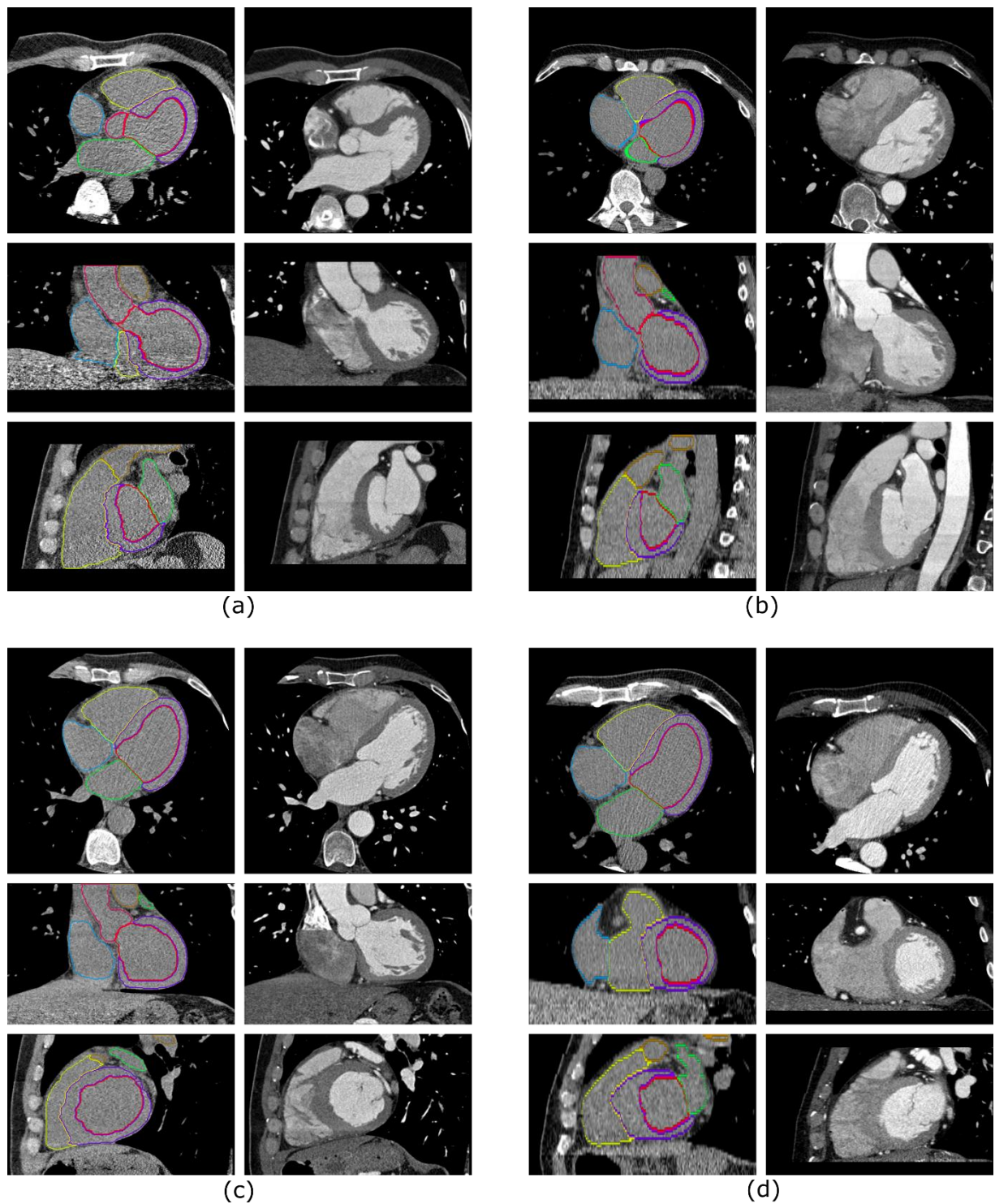

(a)  (b)

(c)  (d)

*Figure 6: Automatic segmentations on axial, coronal, and sagittal planes of non-contrast-enhanced CT (NCCT) images from the single-energy set with coronary CT angiography (CCTA) as a visual reference. (a), (c): Very accurate (grade 1) segmentations on a two thin-slice images from routine clinical care. (b), (d): Very accurate segmentations on two thick-slice images from the orCaScore[27] benchmark. Window/level values on CCTA have been adjusted for similar visual impression.*

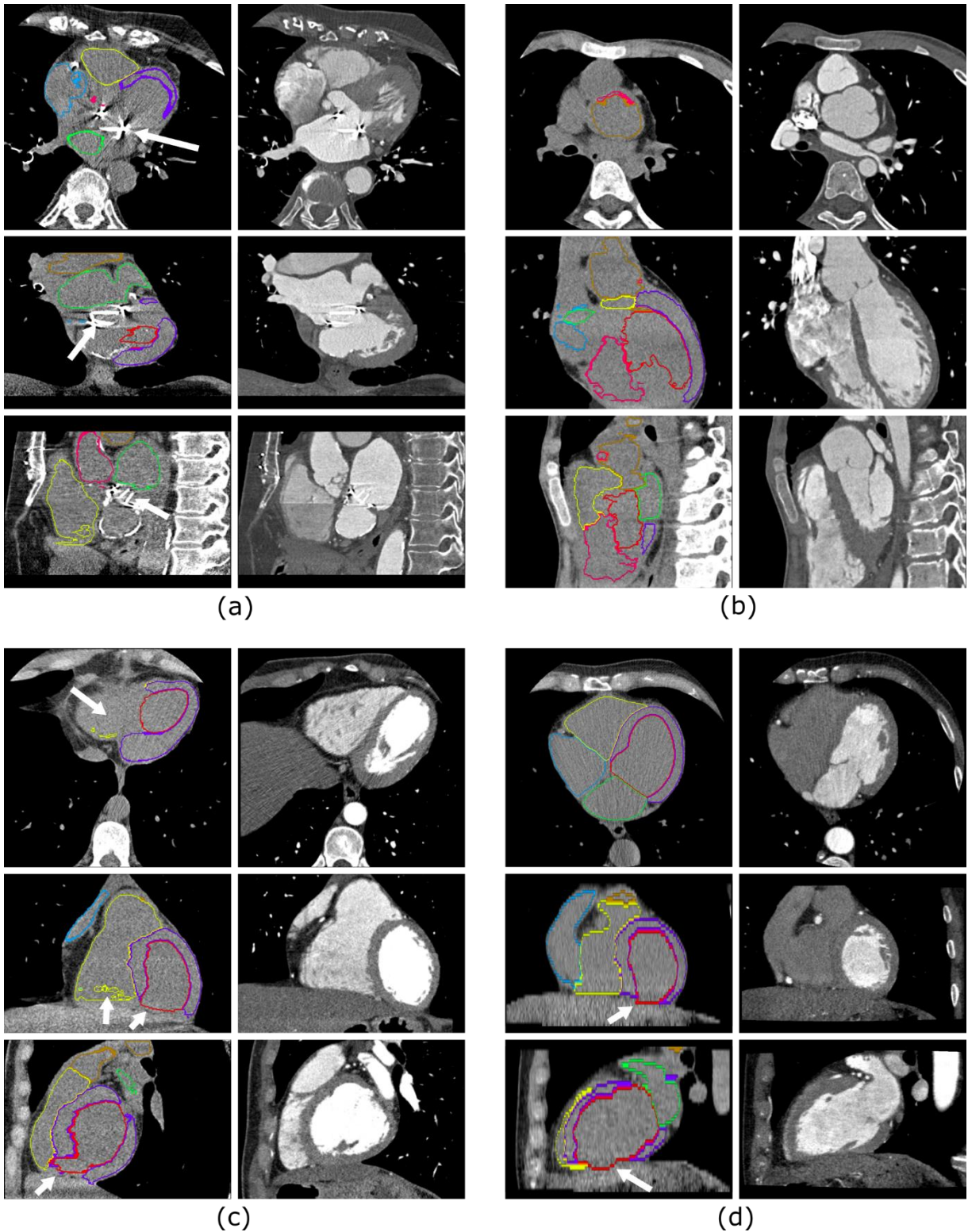

*Figure 7: Automatic segmentations on axial, coronal, and sagittal planes of non-contrast-enhanced CT (NCCT) images from the single-energy set with coronary CT angiography (CCTA) as a visual reference. An incorrect (grade 4) segmentation due to a prosthetic mitral valve (a), and a failed segmentation (grade 5) due to an anatomical variation (b). Incorrect (grade 4) segmentations on a thin-slice image from routine clinical care (c) and on a thick-slice image from the orCaScore[27] benchmark (d).*

# Discussion

We have presented an automatic method for whole-heart segmentation in NCCT using VNC images of a dual-layer detector CT scanner and evaluated this method in a multi-vendor, multi-center study including both thin- and thick-slice images. In an independent set of single-energy NCCT images, we found that the method accurately segmented the cardiac chambers and great vessels in 74% of the cases.

The qualitative results on real NCCT images in the single-energy set show that our method generalizes well to images acquired on single-energy CT scanners from different vendors at different centers, including thin- and thick-slice images. No significant differences were found in mean grades between centers, vendors, and thin- and thick-slice images. While the inter-observer agreement on the fine-grained grade level was moderate due to the subjective scoring system, 74% of the automatic segmentations on NCCT images were rated by both observers as mostly accurate or better. Note that this is a representative and consecutive data set of clinically acquired NCCT images and that no images were excluded. Images on which our method failed to segment most structures accurately (grade 4 or 5) were often affected by artifacts due to cardiac devices or prosthetic heart valves or by anatomical variations such as transposition of the great arteries. Such images were underrepresented in the training set. Given that current learning-based methods have difficulties performing the task on data with previously unseen characteristics, it can be expected that the method underperforms in these cases. Nevertheless, this highlights cases that could be added to the training set to further improve the generalizability of our method. Alternatively, such cases could be rejected for automatic segmentation based on information extracted from electronic health records or possibily an automatic method performing quality control prior to analysis. Observers found that the most challenging structures to segment in the NCCT images were the right ventricle, LV myocardium, and LV cavity. These structures were particularly difficult to segment around the apex. In regions where the myocardium was extremely thin, the method sometimes produced small holes in the myocardium so that the LV cavity and lungs were not always separated by myocardium. It should be noted, however, that we did not enforce any anatomical constraints and that this has little impact on the obtained volumes. Compared with CCTA, or to some extent VNC images, the borders between these structures are hardly distinguishable in NCCT images. Conversely, the outer borders of the cardiac chambers and great vessels are more clearly visible in most parts of the image and therefore, seem to be easier to segment automatically.

We performed a quantitative evaluation of the segmentation performance in VNC using the reference segmentations in CCTA. This highlighted that our method achieved excellent DSC, comparable with related works on whole-heart segmentation in NCCT. Moreover, we found that derived volumes of the cardiac chambers and LV myocardium in VNC closely matched reference volumes in CCTA, despite the fact that contrast enhancement was suppressed in VNC images. However, it should be noted that we quantitatively evaluated segmentation performance on VNC images and not on NCCT images and that there are differences in appearance between VNC and NCCT images. Therefore, conclusions drawn from the quantitative analysis on VNC images cannot be transferred directly to the NCCT domain.

In NCCT images, we found that automatically obtained volumes of all segmented structures were significantly lower than those extracted from the reference standard in CCTA. This effect is not fully explained by segmentation errors, as we even observed volume underestimation in images that were graded as segmented accurately by one of the observers. We hypothesize that structural volume underestimation in NCCT may be due to two factors. First, the presence or absence of contrast agent might influence the interpretation of the position of the outer borders of the heart due to partial-volume effects. A slight shift in the position of this border can have a large impact on volume calculation: e.g., shifting the border of the LV myocardium one voxel inwards can reduce its volume by as much as 9%. A similar effect of underestimating volumes in NCCT with respect to CCTA has previously been found for pulmonary nodule volumes[32]. Second, physiological differences in heart rate, blood pressure, and respiratory state between acquisitions with and without contrast enhancement might affect the volume of individual chambers and the whole heart. This could partly be caused by the administration of nitroglycerine before the CCTA acquisition. This extension effect due to nitroglycerine administration has already been shown for coronary arteries[33] and might be similar for the heart chambers. However, even when absolute volumes differed between CCTA and NCCT images, relative measures like the RV/LV or the RA/LA ratio could still be compared with normal values[34].

Our results show that an automatic segmentation method trained on VNC images generalizes well to NCCT images. Hence, the domain gap between CCTA images and NCCT images is to a large extent bridged by using VNC images. Previous work showed that the volume of contrast agent in CCTA images is replaced by the same volume of blood to generate VNC images[35]. In our study, this is quantitatively supported by the equivalence test which, in line with the literature[24,25], shows that mean attenuation values are equivalent in VNC and NCCT. Nevertheless, differences in image appearance between VNC and NCCT remain and might have been learned by our method. In particular, noise levels are higher in NCCT, and residual contrast patterns can be identified in VNC, as can be seen in Figure 1. To address the former issue, we have applied a smoothing pre-processing step to all images. However, the latter issue is an effect of limitations of current VNC reconstruction algorithms. We expect that as this technology develops, realism of VNC images will improve, and thus similarity to real NCCT images will increase. This further reduction of the domain gap between VNC and NCCT images will likely be reflected in even better generalization between the two domains, and might lead to additional applications for our proposed approach.

In this study, we have shown how data that is acquired as a by-product of image acquisition using a dual-energy CT scanner can be used to train a model that can be applied to any cardiac NCCT image acquired with a range of widely available conventional single-energy CT scanners. While NCCT is unlikely to replace diagnostic CCTA imaging, the proposed method could be used to quantify additional cardiac measures from NCCT images that are acquired for other purposes, such as coronary calcium quantification. For example, the left ventricular mass and geometry could be used for risk prediction of sudden cardiac death[1] or ischemic stroke[2], the left atrial size is known to predict ischemic stroke[3] as well as cardiovascular risk and disease burden[4], and the right atrial volume is an independent risk factor for right-sided systolic dysfunction[5]. In future work, we will investigate to what extent the method is generalizable to routinely acquired non-cardiac NCCT images without ECG synchronization, obtained for other purposes such as lung cancer screening or radiotherapy treatment planning, or other NCCT images in which the heart is visible.

## Conclusion

We have developed a fully automatic deep learning-based method trained on virtual non-contrast images of a dual-layer detector CT scanner to segment the cardiac chambers and great vessels in non-contrast-enhanced CT images. Despite visual differences between virtual non-contrast and real non-contrast CT images, our method produces accurate segmentations and generalizes well to CT images of different vendors from different centers.

## References

1. Narayanan K, Reinier K, Teodorescu C et al. Left ventricular diameter and risk stratification for sudden cardiac death. *J Am Heart Assoc* 2014;3:e001193. doi: 10.1161/JAHA.114.001193.

2. Di Tullio MR, Zwas DR, Sacco RL, Sciacca RR, Homma S. Left ventricular mass and geometry and the risk of ischemic stroke. *Stroke* 2003;34:2380-2384. doi: 10.1161/01.STR.0000089680.77236.60.

3. Nagarajarao HS, Penman AD, Taylor HA et al. The predictive value of left atrial size for incident ischemic stroke and all-cause mortality in African Americans. *Stroke* 2008;39:2701-2706. doi: 10.1161/STROKEAHA.108.515221.

4. Tsang TSM, Barnes MEB, Gersh BJ, Bailey KR, Seward JB. Left atrial volume as a morphophysiologic expression of left ventricular diastolic dysfunction and relation to cardiovascular risk burden. *Am J Cardiol* 2002;90:1284-1289. doi: 10.1016/S0002-9149(02)02864-3.

5. Sallach JA, Tang WHW, Borowski AG et al. Right atrial volume index in chronic systolic heart failure and prognosis. *JACC Cardiovasc Imaging* 2009;2:527-534. doi: 10.1016/j.jcmg.2009.01.012.

6. Sugeng L, Mor-Avi V, Weinert L et al. Quantitative assessment of left ventricular size and function: side-by-side comparison of real-time three-dimensional echocardiography and computed tomography with magnetic resonance reference. *Circulation* 2006;114:654-661. doi: 10.1161/CIRCULATIONAHA.106.626143.

7. Zhuang X, Bai W, Song J et al. Multiatlas whole heart segmentation of CT data using conditional entropy for atlas ranking and selection. *Med Phys* 2015;42:3822-3833. doi: 10.1118/1.4921366.

8. Zheng Y, Barbu A, Georgescu B, Scheuering M, Comaniciu D. Four-chamber heart modelling and automatic segmentation for 3-D cardiac CT volumes using marginal space learning and steerable features. *IEEE Trans Med Imaging* 2008;27:1668-1681. doi: 10.1109/TMI.2008.2004421.

9. Ecabert O, Peters J, Schramm H et al. Automatic model-based segmentation of the heart in CT images. *IEEE Trans Med Imaging* 2008;27:1189-1201. doi: 10.1109/TMI.2008.918330.

10. Zhuang X, Li L, Payer C et al. Evaluation of algorithms for multi-modality whole heart segmentation: an open-access grand challenge. *Med Imag Anal* 2019;101537. doi: 10.1016/j.media.2019.101537.

11. Dey D, Schepis T, Marwan M, Slomka PJ, Berman DS, Achenbach S. Automated three-dimensional quantification of noncalcified coronary plaque from coronary CT angiography: comparison with intravascular US. *Radiology* 2010;257:516-522. doi: 10.1148/radiol.10100681.

12. Hecht HS. Coronary artery calcium scanning: past, present, and future. *JACC Cardiovasc Imaging* 2015;8:579-596. doi: 10.1016/j.jcmg.2015.02.006.

13. Tay SY, Chang P, Lao WT, Lin YC, Chung Y, Chan WP. The proper use of coronary calcium score and coronary computed tomography angiography for screening asymptomatic patients with cardiovascular risk factors. *Sci Rep* 2017;7:17653. doi: 10.1038/s41598-017-17655-w.

14. Lartaud PJ, Rouchaud A, Rouet JM, Nempont O, Boussel L. Spectral CT Based Training Dataset Generation and Augmentation for Conventional CT Vascular Segmentation. *Med Image Comput Comput Assist Interv* 2019. doi: 10.1007/978-3-030-32245-8_85.

15. Noothout JMH, de Vos BD, Wolterink JM, Išgum I. Automatic segmentation of thoracic aorta segments in low-dose chest CT. *Proc. SPIE 10574, Medical Imaging 2018: Image Processing*. doi: 10.1117/12.2293114.

16. Kurugol S, San Jose Estepar R, Ross J, Washko GR. Aorta segmentation with a 3D level set approach and quantification of aortic calcifications in non-contrast chest CT. *Conf Proc IEEE Med Biol Soc* 2012;2012:2343-2346. doi: 10.1109/EMBC.2012.6346433.

17. Išgum I, Staring M, Rutten A, Prokop M, Viergever MA, van Ginneken B. Multi-atlas-based segmentation with local decision fusion – application to cardiac and aortic segmentation in CT scans. *IEEE Trans Med Imaging* 2009;28:1000-1010. doi: 10.1109/TMI.2008.2011480.

18. Kaderka R, Gillespie EF, Mundt RC et al. Cardiac substructure segmentation with deep learning for improved cardiac sparing. *Radiother Oncol* 2019,131:215-220. doi: 10.1016/j.radonc.2018.07.013.

19. Finnegan R, Dowling J, Koh E et al. Feasibility of multi-atlas cardiac segmentation from thoracic planning CT in a probabilistic framework. *Phys Med Biol* 2019;085006. doi: 10.1088/1361-6560/ab0ea6.

20. Zhou R, Liao Z, Pan T et al. Cardiac atlas development and validation for automatic segmentation of cardiac substructures. *Radiother Oncol* 2017;122:66-71. doi: 10.1016/j.radonc.2016.11.016.

21. Haq R, Hotca A, Apte A, Rimner A, Deasy JO, Thor M. Cardio-Pulmonary Substructure Segmentation of CT Images Using Convolutional Neural Networks. *Lecture Notes in Computer Science* 2019;11850:162-169. doi: 10.1007/978-3-030-32486-5_20.

22. Shahzad R, Bos D, Budde RPJ et al. Automatic segmentation and quantification of the cardiac structures from non-contrast-enhanced cardiac CT scans. *Phys Med Biol* 2017;62:3798-3813. doi: 10.1088/1361-6560/aa63cb.

23. Morris ED, Ghanem AI, Dong M, Pantelic MV, Walker EM, Glide-Hurst CK. Cardiac substructure segmentation with deep learning for improved cardiac sparing. *Med Phys* 2019. doi: 10.1002/mp.13940.

24. Ananthakrishnan L, Rajiah P, Ahn R et al. Spectral detector CT-derived virtual non-contrast images: comparison of attenuation values with unenhanced CT. *Abdominal Radiology* 2017;42:70-709. doi: 10.1007/s00261-016-1036-9.

25. Javadi S, Elsherif S, Bhosale P et al. Quantitative attenuation accuracy of virtual non-enhanced imaging compared to that of true non-enhanced imaging on dual-source dual-

energy CT. *Abdominal Radiology* 2020 [published online February 12, 2020]. doi: 10.1007/s00261-020-02415-8.

26. Van Hamersvelt RW, Išgum I, de Jong PA et al. Application of speCtral computed tomogrAphy to impRove specIficity of cardiac compuTed tomographY (CLARITY study): rational and design. *BMJ Open* 2019;9:e025793. doi: 10.1136/bmjopen-2018-025793.

27. Wolterink JM, Leiner T, de Vos BD et al. An evaluation of automatic coronary artery calcium scoring methods with cardiac CT using the orCaScore framework. *Med Phys* 2016;43:2361-2373. doi: 10.1118/1.4945696.

28. Bruns S, Wolterink JM, van Hamersvelt RW, Zreik M, Leiner T, Išgum I. Improving myocardium segmentation in cardiac CT angiography using spectral information. *Proc. SPIE 10949, Medical Imaging 2019: Image Processing*. doi: 10.1117/12.2512229.

29. Johnson J, Alahi A, Fei-Fei L. Perceptual losses for real-time style transfer and super-resolution. *Comput Vis ECCV* 2016. doi: 10.1007/978-3-319-46475-6_43.

30. Milletari F, Navab N, Ahmadi S. V-net: Fully convolutional neural networks for volumetric medical image segmentation. *Proc Int Conf 3D Vis* 2013. doi: 10.1109/3DV.2016.79.

31. Abadi S, Roguin A, Engel A, Lessick J. Feasibility of automatic assessment of four-chamber cardiac function with MDCT: Initial clinical application and validation. *Eur Radiol* 2009;74:175-181. doi: 10.1016/j.ejrad.2009.01.035.

32. Den Harder AM, Bangert F, van Hamersvelt RW et al. The Effects of Iodine Attenuation on Pulmonary Nodule Volumetry using Novel Dual-Layer Computed Tomography Reconstructions. *Eur Radiol* 2017;27:5244-5251. doi: 10.1007/s00330-017-4938-1.

33. Dewey M, Hoffmann H, Hamm B. Multislice CT Coronary Angiography: Effect of Sublingual Nitroglycerine on the Diameter of Coronary Arteries. *Rofo* 2006;178:600-604. doi: 10.1055/s-2006-926755.

34. Fuchs A, Mejdahl MR, Kühl T et al. Normal values of left ventricular mass and cardiac chamber volumes assessed by 320-detector computed tomography angiography in the Copenhagen General Population Study. *Eur Heart J Cardiovasc Imaging* 2016;17:1009-1017. doi: :10.1093/ehjci/jev337.

35. Mendonça PRS, Lamb P, Sahani DV. A Flexible Method for Multi-Material Decomposition of Dual-Energy CT Images. *IEEE Trans Med Imaging* 2014;33:99-116. doi: 10.1109/TMI.2013.2281719.

## Acknowledgements

This study was funded by the Dutch Technology Foundation (STW, perspectief, P15-26) with participation of Philips Healthcare, Haifa, Israel.

## Conflicts of Interest Statement

Ivana Išgum and Tim Leiner received institutional research projects by the Dutch Technology Foundation co-funded by Pie Medical Imaging and Philips Healthcare (P15-26), the Netherlands Organisation for Health Research and Development with participation of Pie Medical Imaging (104003009). Ivana Išgum received an institutional research project by the Dutch Technology Foundation co-funded by Pie Medical Imaging (12726). Ivana Išgum and Tim Leiner are cofounders and shareholders of Quantib-U BV, Utrecht, the Netherlands. The authors declare that the research was conducted in the absence of any additional commercial or financial relationships that could be construed as a potential conflict of interest.